\documentstyle[12pt,preprint,epsf]{aastex}
\def\omegacen{$\omega$-Centauri~}
\def\msat{\mu}

\def\kpc{\, {\rm kpc} }
\def\msun{M_\odot}
\def\lsun{L_\odot}
\def\kms{\, {\rm km \, s}^{-1} }
\def\BF{\mathbf}
\def\bey{\begin{eqnarray}}
\def\eey{\end{eqnarray}}
\def\beq{\begin{equation}}
\def\eeq{\end{equation}}

\def\edcomment#1{\iffalse\marginpar{\raggedright\sl#1\/}\else\relax\fi}
\reversemarginpar

\begin{document}
\title{Can satellites deliver substructures and black holes to inner halo by dynamical friction?}
\author{HongSheng Zhao\footnote{PPARC Advanced Fellowship (hz4@st-andrews.ac.uk)}}
\affil{Institute of Astronomy, Cambridge, CB30HA, UK}
\affil{University of St. Andrews, School of Physics and Astronomy, KY16 9SS, Scotland}
\begin{abstract}
A fully analytical formulation is developed to make dynamical friction
modeling more realistic. The rate for a satellite to decay its orbit in a 
host galaxy halo is often severely overestimated when applying the ChandraSekhar's 
formula without correcting for the tidal loss of the satellite and the
adiabactic growth of the host galaxy potential over the Hubble time. As a
satellite decays to the inner and denser region of the host galaxy, the high
ambient density speeds up their exchange of energy and angular momentum, but 
shrinks the Roche lobe of the satellite by tides. Eventually both processes of orbital 
decay and tidal stripping hang up altogether once the satellite is light enough. 
These competing processes can be modeled analytically for a satellite 
if we parametrize the massloss history by an empirical formula. 
We also incorporate the adiabatic contraction of orbits due to growth of the 
host potential well. Observed dwarf galaxies often show a finite density core,
which determines how much inwards its remnants could be delivered to the host
galaxy. Ghost streams or remnant cores or globular clusters of satellites
should populate preferentially the outer halo (e.g., the Magellanic stream and
the Ursa Minor dwarf), rather than the inner halo (e.g., the Sgr stream and the
Omega-cen cluster). Massloss due to strong tides in the inner galaxy also makes
it problematic for any small central black holes in nuclei of satellite galaxies 
to decay in orbit and merge into the center of the host galaxy.
\end{abstract}

\section{Introduction}

Current theory of galaxy formation favors the idea that galaxies form hierarchically
by merging smaller lumps or satellites.  Dynamical friction dissipates the orbital energy of 
the satellites so that they sink deep into the host galaxy potential well, where they are
disintegrated and virialized via baryonic feedbacks and tidal stripping.  These processes might have
determined the density profile of virialized halo of the host galaxy (Syer \& White 1998, Dekel et al. 2003).

There are about 150 and 300 globular clusters, and a few dozen dwarf satellites
of mass $10^{6-9}\msun$ in the Milky Way and M31 respectively.
It is tempting to associate 
dwarfs satellites and globular clusters in the Milky Way as the markers/remnants of past 
hierarchical merging events.
Indeed there are several examples of possible streams of remnants in the Milky Way 
(Lynden-Bell \& Lynden-Bell 1995) including the recently found Galactic ring or Carnis Major dwarf galaxy,
traced by a grouping of globular clusters (Martin et al. 2003).  
A giant stream is found in the Andromeda galaxy (McConnichie et al. 2003, Ferguson et al, 2002).
Among these the Sagittarius dwarf galaxy stream (Ibata et al. 1997) between radius 10-50 kpc 
from the Galactic center is perhaps the best example.  
It brings in at least 5 globular clusters to the inner halo, including M54, 
the 2nd most massive cluster of the Milky Way.   
A still mysterious object is \omegacen, with about $10^6\lsun$ at 5 kpc from the Galactic center
which has the morphology of a globular cluster, but has multiple epoches
of star formation and chemical enrichment (see Gnedin et al. 2002 and 
the \omegacen symposium).  Another example is the G1 cluster, 
a very massive globular-like object with about $10^6\lsun$ at about 40 kpc
from the M31 center (Meylan et al. 2001). A system (NGC1023-13) almost
identical to G1 is also found in the S0 galaxy NGC1023, at a
projected distance of about 40 kpc from the host galaxy (Larsen 2001).  
Freeman (1993) suggested that such systems are the remnants of nucleated dwarf satellites,
with their outer tenuous dark matter and stars being removed by galaxy tides.  
M31 also has an unusual collection of 
clusters as luminous as \omegacen within 5 kpc in projection from its double-peaked center.
Given the above evidences or signs for infalling objects in our galaxy and M31,
it is interesting to ask whether some of the inner globulars could have
also been the result of mergers.  Beyond the Local Group,
minor mergers are sometimes speculated as the mechanism to deliver massive
black holes and gas material into the nucleus of an AGN to account
for the directions of the jets and nuclear dusty disks.  A very interesting
related issue is whether giant black holes acquire part of the mass
by merging the smaller black holes in the nuclei of infalling satellites.

For this paper we revisit the basic theoretical questions: what is 
the condition for a dwarf galaxy to decay into the inner halo? 
What are the possible outcomes of tidal stripping of a dwarf satellite? 
How often do we get a system like \omegacen or a naked black hole near the host center?
The answers to these questions will help us to test the validity of the theory of hierarchical
merger formation of galaxies.  
The key mechanism for satellites to enter the inner galaxy
is dynamical friction, where the gravity of the satellite creates a wake 
of overdensities in the particle distribution of the host galaxy,
which in turn drags the motion of the satellite with a force
proportional to $m(t)^2$, where $m(t)$ is the mass of the satellite.  
Another process is tidal disruption, where the object sheds mass with each 
pericenter passage, and the remnants are littered along the orbit of the satellite.  
The above two processes compete with and regulate each other:
orbital decay increases the tidal field, which reduces the mass of the satellite, 
hence slows down the orbital decay.  Some examples of these effects
have been shown in Zhao (2002) in the case of \omegacen.

The analytical formula of ChandraSekhar  is widely used for 
gaining insights on dynamical friction because of the time-consuming nature
the more rigorous N-body numerical simulation approach.
It is a customary practice in previous works to model the orbital decay of 
a satellite as a point mass of a fixed mass.  However,  
{\it the fixed mass approximation is unnecessary}
and could seriously overestimate dynamical friction because of neglecting 
massloss ${dm(t) \over dt}$.  
It is essential in calculations of satellite orbits to model 
the dynamical friction and massloss together since they regulate each other.

In the past the massloss and the orbital decay are often modeled in the ab. initial fashion, 
resulting coupled non-linear equations without simple analytical solutions.  
In such models,  
the satellite mass is often modeled as a function of the satellite's tidal radius,
hence various factors come in, including the orbital position of the satellite,
the density profile of the satellite
(e.g., Jiang \& Binney 2000, Zhao 2002, Mouri \& Taniguchi 2003, Kendall et al. 2003).
However, these complications are not always necessary since the massloss history is 
rather similar in simulations with very different initial conditions
(the mass is generally a stair-case like a function of the time), so 
could be parametrized in an empirical fashion, by-passing the uncertain assumptions of the satellite
initial profile.  This could be useful for exploring a large parameter space of the satellite
initial condition.

Another unnecessary approximation but common practice is to use a static potential for the host galaxy.
This again is unphysical since galaxy halos do grow in hierarchical formation scenario
partly because of galaxy merging, and partly because of the adiabatic contraction
of the baryonic disk and bulge; galaxy rotation curves $V_{\rm cir}(r,t)$ 
can change by significantly before and after the formation of baryonic disks and bulges in mass models for the 
Milky Way and M31 (Klypin, Zhao \& Somerville 2002) and in generic CDM simulations with baryons (e.g. Wright 2003).
The growing gravitational force tends to restrain the radial
excursions of the satellite while preserving the angular momentum.  
The dynamical friction or drag force 
is also proportional to the growing density $\rho(r,t)$ of ambient stars and dark particles
in the host galaxy. 

In fact, it is conceptually simple to incorporate massloss and growing potential 
while keeping the problem 
analytically tractable: the deceleration $-{dv \over dt}$ is simply proportional to the 
satellite mass $\msat(t)$ and the ambient density $\rho(r,t)$ of the host galaxy at the time $t$.  
Without massloss ChandraSekhar's formula would predict very efficient braking
of the orbits, enough to make a high-mass satellite of $\sim 10^{10}\msun$
(the mass of the LMC or M33 sized object) to decay from a circular
orbit at $\sim 100$kpc to the very center of 
a high brightness galaxy in a Hubble time, delivering remnants into the inner galaxy.
Here we study the effect of massloss and growing potential on the 
result of orbital decay, and the distribution of remnants.  
We present fully analytical results for calculating the decay rate for satellites
on eccentric orbits in a scale-free growing isothermal potential.

The structure of the paper is following: in S2 we give the analytical formulation
of the problem, in S3 and S4 we present results of    
application of our analytical model to globulars and dwarf satellites, 
in S5 we discuss and in S6 we summarize.

\section{Analytical model of orbital decay of a shrinking satellite in a growing host halo}

Consider a satellite moving with a velocity ${\BF v_s}$ on a rosette-like orbit 
in a spherical host galaxy potential $\phi(r,t)$ with a rotation curve $V_{\rm cir}$.  
Assume it has an initial mass $m_i$, 
the orbital decay from an initial time $t=t_i$ to the present day $t=t_0$
can be modeled using ChandraSekhar's dynamical friction formula (Binney \& Tremaine 1987)
\beq\label{dv}
{d{\BF v_s} \over dt} = - {{\BF v_s} \over t_{\rm frc}} - {V_{\rm cir}^2 \over r^2}{\BF r} 
\eeq
where $t_{\rm frc}$ is the instantaneous dynamical friction time.
Manipulating the equation, we find 
the (specific) angular momentum $j(t)$ of the satellite decays as
\beq
{d j(t) \over dt} = {\BF r} \times {d{\BF v_s} \over dt} = - {j(t) \over t_{\rm frc}} , 
\qquad j(t)= {\BF r}(t) \times {\BF v_s}(t).
\eeq
Here the dynamical friction time is given by
\beq
t_{\rm frc}^{-1} = 
\left[4\pi G\rho(r,t)\right] {G m(t) \over V_{\rm cir}^3} \xi, 
\eeq
where $m(t)$ is the mass of the satellite, 
$\rho(r,t)$ is the density of host galaxy at the orbital radius $r(t)$,
and $\xi$ is some dimensionless function 
of the rescaled satellite speed $\left({|{\BF v_s}| \over V_{\rm cir}}\right)$.

Here we adopt 
a Singular Isothermal Spherical (SIS) host galaxy model with a rotation curve generally growing with time
$V_{\rm cir}(t)$; it is normalized by the present time value $V_{\rm cir}(t_0)=V_0$.  
In this model the potential and density are given by
\beq
\phi(r,t)=V^2_{\rm cir}(t) \ln r,\qquad
\rho(r,t) = {V^2_{\rm cir}(t) \over 4 \pi G r^2}.
\eeq
Since the stars and dark matter velocity distribution in the host galaxy 
is an isotropic Gaussian, the dimensionless velocity function
\beq
\xi(u) = \left[{\rm erf}(u)-{2u \over \sqrt{\pi}} \exp\left(-u^2\right)\right] \ln\Lambda,
\qquad u(t) \equiv \left({|{\BF v_s}| \over V_{\rm cir}}\right),
\eeq
where the dimensionless Coulomb logarithm $\ln \Lambda$ is typically between unity and ten.

Combining the above equations we find the decay rate for the angular momentum is given by
\beq
{d j(t) \over dt} = - {j(t) \over t_{\rm frc}}, \qquad  
{1 \over t_{\rm frc}} = { G \msat(t) V_{\rm cir}(t) \over j(t)^2},
\eeq
where we define an effective bound mass of the satellite at time $t$ 
\beq
\msat(t) \equiv m(t) \xi(u) u^2 \cos^2(\alpha), \qquad u(t)={|{\BF v_s}(t)| \over V_{\rm cir}(t)},
\eeq 
where $u(t)$ and $\alpha(t)$ are the rescaled speed and the pitch angle of the orbit
at time $t$.  As we can see, the effective mass $\msat(t)$ lumps together several time-varying 
factors, including the satellite massloss and orbital epicyles.  Introducing this
{\it effective} mass simplifies the physics details because 
the braking rate of specific angular momentum $-dj(t)/dt$ is now simply proportional to $\msat(t)$.

Integrating over time, we find the change in the specific angular momentum between
time $t_i$ to $t_0$ is given by 
\beq
-\Delta j^2 =  j^2(t_i) - j^2(t_0) = 2  G m_i V_0 \tau, 
\qquad \tau \equiv (t_i-t_0) \gamma \equiv \int_{t_i}^{t_0}  {\msat(t) V_{\rm cir}(t) \over m_i V_0} dt,
\eeq
where we define $\gamma$ as the reduction factor from fixed satellite model, and define 
$\tau$ as the effective duration of dynamical friction,
$m_i$ and $V_0$ are the mass at initial time $t_i$ and the rotation curve at the present $t_0$.  
The above formula allows us to calculate
the evolution of the specific angular momentum $j(t)$ of a generally
{\it eccentric satellite orbit}.

It is interesting to ask
what the necessary condition is for an outer halo satellite to reach the inner galaxy
or even the center.  These satellites will be destroyed with their remnants 
in the inner galaxy, e.g., by the disk and/or bulge shocking.  
To facilitate the comparison with observations,
it is sometimes better to define an instantaneous orbital size
\footnote{We can make this defination irrespective of the eccentricity of the orbit, which we does not
enter our calculation explicitly.  Roughly speaking the orbital size $S(t)$ is the geometrical mean of
the pericenter radius and apocenter radius in a static potential.} 
\beq
S(t) \equiv {j(t) \over V_0}
\eeq
at time $t$.  Let the present orbital size 
$S_0={j_0 \over V_0} \le R_{\rm disk}\sim 15$kpc,
and define $j_{\rm disk}=R_{\rm disk}\times V_0$ as the critical angular momentum to reach
within the disk at the present time.  Manipulating eq. (8)
we find the initial angular momentum $j_i$ or orbital distance $S_i={j_i \over V_0}$ must satisfy
\beq
S_i =
\left[ S_0^2 + {2G m_i \tau \over V_0} \right]^{1 \over 2}, \qquad 0\le S_0 \le R_{\rm disk}, 
\qquad \tau \equiv \int_{t_i}^{t_0}  {\msat(t) V_{\rm cir}(t) \over m_i V_0} dt.
\eeq
In  other words, satellites initially on orbits larger than $S_i$ would not be able to deliver
a globular or dwarf galaxy to the inner galaxy irrespective of mass loss assumed.

Eq. (8) is our main analytical result.  There is also an interesting
geometrical interpretation to it.  Consider a satellite of mass $m(t)$ on 
a circular orbit in a static SIS potential,  we have
\beq
\ln \Lambda \approx 2.5, \qquad \msat(t)=m(t)\xi(1)=0.42 \ln \Lambda m(t) \approx m(t),
\eeq 
where we took a typical value for the Coulomb logarithm (e.g., Panarrubia, Kroupa \& Boily 2002).
We can rewrite the angular momentum equation (eq. 8) as
\beq
\left| \pi \Delta S^2 \right| \approx  
{2\pi G \left<m\right> \over V_0^2} \left(  V_0 \Delta t\right), \qquad
\left<m\right> = {\int_{t_i}^{t_0} m(t) dt \over t_0-t_i} 
\eeq
where the l.h.s. is the area swepted by the decaying orbit, 
and in the r.h.s.  
$2\pi G \left<m\right> V_0^{-2}$ is the "circumference of influence" of the satellite,
and the $V_0\Delta t=V_0(t_0-t_i)$ is the length of the satellite's orbital path,
and the multiplication of the two is the area swept by the circumference of influence
of the satellite.  The above equation implies that the two areas are comparable.
Interestingly the decay of the orbits depends on the satellite mass  
through the approximation $\Delta S^2 \propto \left<m\right> \Delta t $,
i.e., it is the average mass of the satellite that determines the rate
of orbital decay.   Finally note eq.(8) applies to the evolution of specific angular momentum 
(S.A.M.) of an eccentric orbit in a time-varying potential. 
A complete description of the orbits should also include the evolution of the orbital energy,
which unfortunately is more complex analytically, and is not studied in detail here.
It is not essential for our conclusion, but is perhaps convienent to assume
efficient dynamical decay of the orbital energy 
during the pericentric passages hence the radial motion is damped and 
orbit circularizes at the end.  Adiabatic growth of the potential also tends to circularize
the orbits.

\subsection{Models for a growing host potential and a diminishing satellite}

Classical singular isothermal model assumes a fixed potential or a time-invariant rotation curve.
In reality galaxies grow substantially from redshift of a few to now.  
The growth of the dark halo is scale-free in the hierarchical scenario,  
so it is plausible to approximate the growth of the rotation curve
as a power-law in time as follows,
\beq
V_{\rm cir}(t)  = V_0 \left({t \over t_0}\right)^{p},\qquad 0<t<t_0,
\eeq
where the rotation speed at some earlier time $0<t<t_0$ is generally smaller than 
the present day ($t=t_0=14$Gyr) value $V_0$.    
At the present epoch $V_{\rm cir}(t_0)=V_0=200\kms$ is appropriate for the Milky Way.
A few sample evolution models are shown in Fig.1.  A value of ${1 \over 9} \le p \le {1 \over 3}$ implies
an evolution of 10\%-25\% in rotation speed from redshift one to now, which seems 
reasonable in models of the Milky Way before and after the formation of 
the disk and the bulge (Klypin, Zhao, Somerville 2001; Wright 2003).

Previously the tidal stripping of the satellite is either modeled
by N-body simulations (e.g., Tsuchiya et al. 2003, Johnston et al. 1999)
or semi-analytically using tidal radius criteria (e.g., 
Zhao 2002, Kendall, Magorrian \& Pringle 2003). 
\footnote{Recently the latter approach has also been extended to model the orbital decay and
evaporation-induced massloss in the dense star cluster near the Galactic center
(Mouri \& Taniguchi 2003, McMillan \& Portegies-Zwart 2003). }
In both approaches one needs to assume
a rigorous description of the mass profile of the progenitor satellite.  
It is unclear whether this approach can model efficiently
the realistic large scatter in the circular velocity curves of observed dwarf galaxies
(see, e.g., Fig.4a).

Here we take a very different approach.
We model the mass of the progenitor galaxy as a simple function of 
time with two free parameters.  We do not need an explicit prescription
of the satellite density profile.  Motivated by the massloss history
typically seen in N-body simulations, 
the satellite mass is modeled to decay in a rate between exponential and linear massloss.
To simplify the calculations, we lump together several uncertain factors,
and simply assume the {\it effective} mass of the satellite is shed following the following recipe,
\beq
\msat(t) = m_i \left[1 - \left(\hat{t}-{\sin 2N\pi \hat{t} \over 2N\pi}\right)\left(1-{m_0^{1 \over n}\over m_i^{1 \over n}}\right) \right]^{n},
\qquad \hat{t}={t-t_i\over t_0-t_i},
\eeq
where 
$m_0$ and $m_i$ are the mass at the present $t=t_0=14$Gyrs and 
when the satellite falls in $t=t_i$, and the quantities
$0 \le \hat{t} \le 1$ and $N\sim 1-100$ 
are the rescaled dimensionless time and the number of pericentric passages
between time $t_0$ and $t_i$.  
The parameter $n$ is tunable
with the $n=1$ model having a constant rate of massloss,
and the $n \rightarrow \infty$ having an exponential massloss;
and $n=-1$ is a roughly power-law decay.
In simulations we typically see somewhere in between these cases:
$0<n \le 1$ for Plummer or King model satellites with 
a sharp fall off after an initially linear massloss (e.g., Panarrubia et al. 2002); for isothermal 
satellite models, the massloss is close to linear $n=1$ (Zhao 2002).  
Note for a completely disrupted satellite $m_0 \sim 0$.   
In N-body simulations satellite losses mass mainly in bursts near pericentric passages,
so the mass is a descending staircase like function of time.  
This is simulated fairly well by our formula.  
A few sample massloss histories are shown in Fig.1 where
we assume $N=10$ pericentric passages from $t_i=4$Gyr to now $t_0=14$Gyr.

Dynamical friction can be greatly reduced by satellite massloss and host growth.
The reduction factor $\gamma$ is computed by substituting eqs. (13-14) into eq. (8).
For a rough estimation of the reduction factor $\gamma$, 
consider a {\it Gedanken} experiment where a satellite 
enters a growing host galaxy at time $t_i=0$ 
right after the big bang and is completely dissolved ($m_0=0$) by the time $t_0$.
Assume the orbits are circularized, so $N \rightarrow \infty$.
We find $\gamma={p!n!\over (p+n+1)!} = {1 \over 6}$ if the satellite 
losses mass linearly with time ($n=1$) in a linearly growing halo ($p=1$). 
This estimate is perhaps to the extreme.  
In reality the formation and mergers of the satellites are probably 
over an extended period of the Hubble time, perhaps 
starting around redshift of 1.5 ($t_i=4$Gyr), ending around now ($t_0=14$Gyr).
Hereafter we consider mostly models with 
$t_i=4$Gyr, and $t_0=14$Gyr, ${1 \over 9} \le p \le {1 \over 3}$ 
and $m_0 \ll m_i$.

\section{Results of application to globulars and dwarf satellites}

Section 2 gives the formalism to predict analytically
the evolution of the angular momentum of the satellite 
for a satellite with any massloss history on an eccentric orbit
around a time-varying potential. 

Clearly not all satellites could reach the galactic center as a globular or a naked massive
black hole.  Dynamical friction is basicly turned off if the satellite bound mass drops 
below $10^9\msun$ before reaching the inner galaxy.
To reach the inner, say, 15 kpc of the host galaxy, which is roughly 
the truncation radius of the outer disk of the Milky Way,
a satellite must have presently a specific angular momentum 
\beq
0 \le j_0 \le j_{\rm disk} \equiv R_{\rm disk}V_0=15{\rm kpc}\times 200 \kms.
\eeq
In comparison the present specific angular momentum of some of the known
satellites are given in Table 1.  The Magellanic stream and Ursa Minor have  
$j_0 \sim 15000\kms$kpc much larger than $j_{\rm disk}$.

\begin{deluxetable}{llll}
\tablecaption{Specific angular momentum (SAM) or orbital size of known satellites of the Local Group \label{tbl-1}}
\tablehead{\colhead{$j_0={\BF r}{\rm kpc} \times {\BF v} \kms$} & \colhead{$S_0={j_0 \over 200\kms}$ (kpc)}& \colhead{Object} &\colhead{Ref}}
\startdata
$5{\rm kpc}\times 50 \kms $   & 1.25 & {\rm \omegacen},
&{Dinescu et al. 1999}\\
$16{\rm kpc}\times 260 \kms $ & 20.8 & {\rm Sgr stream,}
&{Ibata et al. 1997}\\
$60{\rm kpc}\times 250 \kms $ & 75   & {\rm Magellanic stream,}
&{Kroupa \& Bastian 1997}\\
$70 {\rm kpc}\times 200\kms $ & 70   & {\rm Ursa Minor} 
&{Schweizer et al. 1997} \\
$138 {\rm kpc}\times 310\kms $& 213  & {\rm Fornax}
&{Piatek et al. 2002} \\
$150{\rm kpc}\times 20 \kms $ & 15   & {\rm M31 stream\tablenotemark{a}}
&{McConniche et al. 2003} \\
$15{\rm kpc}\times 200 \kms $ & 15   & {\rm Canis Major dwarf galaxy}
&{Martin et al. 2003}
\enddata
\tablenotetext{a}{Assuming NGC205 or M32 is near the pericenter of the stream orbit.}
\end{deluxetable}

Suppose there was a population of hypothetical dwarf satellites 
in the outer halo with a specific angular momentum comparable to that of the orbits of LMC, Ursa Minor and Fornax.  
Suppose their initial effective mass $m_i$ is between Ursa Minor 
and the LMC ($10^{7-10}\msun$) at sometime $t_i$ between $4-14$Gyr. 
We integrate forward in time to answer the question where their remnants are.
Clearly the effect of orbital decay is maximized if we take the longest evolution time (10Gyr),
the highest initial satellite mass and smallest initial orbit $j_i=60{\rm kpc}\times 250\kms$.  
This is the case shown by the hatched region in Fig. 2 
with the satellite starting from the upper right corner and ending to the lower left with a mass $10^7\msun$.  
The vertical axis is a characteristic orbital distance of the satellite,
expressed in terms of the SAM (specific angular momentum) divided by a characteristic velocity $200\kms$.
This orbital distance is roughly the geometrical mean of the apocenter and pericenter.
Models are shown in order of increasing dynamical braking.
The upper shaded zone are models with between
exponential and linear massloss and a moderate evolution of the potential ($\infty \ge n \ge 1, p={1 \over 3}$), 
the lower shaded zone are models with between linear to accelerated massloss and a minor evolution of the potential 
($1>n>0.3,p={1 \over 9}$).  Qualitatively speaking the orbital decay appears to be only modest in all cases,
the remnants are generally not delivered to the inner galaxy.  

The condition for a satellite to deliver a low-mass substructure to the inner halo (cf. eq.15)
or a $10^6\msun$ black hole to the galaxy center 
is summarized in Fig.3.  The satellite
must be within 20kpc for the past Hubble time 
for a low-mass ($10^{7-9}\msun$) progenitor.  It could be at a modest distance
of 40-50kpc if the progenitor was very massive ($10^{10}\msun$) with a linear or accelerated massloss
($n<1$) and little evolution of the galactic potential ($p \ll 1$).
Progenitors of the inner halo substructures or central black hole cannot be 
on orbits of specific angular momentum comparable to the LMC, Ursa Minor or Fornax.
This illustrates the difficulty of making systems such as \omegacen as the nucleus of
a stripped-off dwarf galaxy starting from the very outer halo. 
Likewise it is difficult for a minor merger to bring in a million solar mass black hole
to the host galaxy center.  The progenitor's orbit must be radial and well-aimed 
at the galaxy center such that the progenitor's tangential velocity is  
$\le 1\kms$ if the satellite comes from an initial distance of 1Mpc.

\section{Tidal constraints on internal density profile and circular velocity curve of satellite}

The basic feature of our dynamical friction models is that they 
bypass any information of the internal density profile
of the satellite by specifying $m(t)$ directly.  Nevertheless 
we can infer the internal mass distribution, or internal circular velocity curve,
of the satellite using the tidal radius criteria.  More specificly
the internal circular motion $u_{cir}$ at the tidal radius $r_t$ should be 
in resonance with the satellite's orbital velocity $V(t)$ at the pericenter $r(t)$, 
meaning that their angular frequencies are equal with 
\beq
{u_{cir} \over r_t} = t_{cr}^{-1}  = {k V_{cir} \over r(t) }, 
\eeq
where $t_{cr}$ is the crossing time, the factor $k$ 
is the boosting factor of the pericenter velocity due to eccentricity,
and the velocities
\beq
u_{cir}=\sqrt{G m(t) \over r_t}, \qquad k V_{cir}= {j(t) \over r(t)}.
\eeq
We fix $k=\sqrt(3)$ hereafter, appropriate for a very eccentric orbit with peri-to-apo
radius ratio of $1:5$, e.g., the orbit of the Sgr dwarf galaxy; 
we neglect any circulation of the orbit.    
Use substitutions to eliminate $r(t)$, we have
\beq
{u_{cir}^3 \over G m(t)} = t_{cr}^{-1}  = {k^2V_{cir}^2 \over j(t)},
\eeq
and eliminating $u_{cir}$ we have
\beq
\left(G\rho_t\right)^{1 \over 2} \propto \left({G m(t) \over r_t^3 }\right)^{1 \over 2} = t_{cr}^{-1}
\eeq
where $\rho_t$ is the satellite's overall mean density at tidal radius $r_t$.
This means that as the satellite sinks in to the halo with a shrinking mass $m(t)$ and 
specific angular momentum $j(t)$, the satellite's mean density $\rho_t$ increases with the
ambient density, hence the tidal radius $r_t$ shrinks with the satellite's mass $m(t)$,
like what happens with peeling off an onion.  The tidal peeling-off process effectively
maps out the internal mass radial profile of the satellite, i.e., $m(r_t)$ implicitly.

From the mass radial profile $m(r_t)$ 
we can derive an internal mean density profile $\rho_t(r_t)$ or more 
usefully an internal circular velocity profile $u_{cir}(r_t)$, which is shown in Fig. 3.
Here we consider only satellites sink in a moderately growing ($p=1/3$) host potential,
starting with a mass and angular momentum similar to the present values of the LMC.
Depending on the assumed massloss history $m(t)$, we have a range of implied
circular velocity curves $u_{cir}$ vs $r_t$.  
At the lower end we have models with rapid ($n \gg 1$) or exponential ($n \rightarrow \infty$)
massloss, which show an overall solid-body circular velocity due to 
an overall uniform density profile; such satellites are disrupted rapidly once the 
ambient density reaches the uniform internal density.  
At the higher end we have models with $n \ll 1$, which show a nearly Keplerian
velocity at large radii due to a highly stratified density profile, which behaves 
like a point-mass at large radii; such concentrated satellites are able to retain 
most of their mass as they spirial deep into the denser and denser region of the host galaxy
before eventual disruption.  
The circular velocity curves of all models 
show a solid-body core due to a finite central density;
the faster the rise of the solid-body core, the denser the satellite central density,
and the further inwards the final disruption location of the satellite.
In the middle we have models with linear massloss ($n=1$), which show 
a nearly flat velocity curve at large radius and a solid-body core.
It can be understood analytically why 
singular isothermal satellite losses its mass linearly with time
in a singular isothermal host halo of fixed $V_{cir}$:
From the tidal criteria (eq. 18) and the orbital evolution (eq. 6) we have 
\beq
m(t) \propto j(t), \qquad {dj(t) \over dt} \propto {m(t) \over j(t)}, 
\eeq   
hence we have 
\beq
{dm(t) \over dt} \propto {dj(t) \over dt} = cst,
\eeq
meaning a linear massloss.  

The above formulation is based on the simplifying static picture that the satellite's mass is peeled off
in successive layers at the shrinking tidal radius.   The picture in N-body simulations 
is more complicated, since satellite particles at all radii, e.g., the center of the satellite, 
could be escape at any time.  So in principle eq. (19) sets only a lower limit on
the initial mass profile of the satellite.  We shall proceed without making this correction, although
it is possible to introduce an empirical factor to correct this, 
e.g., as in Jian \& Binney (2000). 

Consider satellites evolving from 
an initial mass $10^{10}\msun$ at time $t_i=4$Gyr
to a final remnant mass $10^6\msun$ at time $t_0=14$Gyr.  
The mass profiles of the satellites can be predicted
depending on the massloss parameter $n$.  Several circular velocity
curves are shown in Fig.4b and Fig.5.
Compare these with the rotation curves of several nearby dwarf galaxies
of virial mass $\sim 10^{10}\msun$ (Fig.4a).  
Observed dwarf galaxies typically have a solid-body core instead
of a cold dark matter cusp.  
It is remarkable that our predicted circular velocity curves
resemble very well 
those of observed dwarf galaxies, for a variety of initial and final
parameters of the satellites and the host galaxy.
The fact that our models give reasonable mass profiles of satellites
justifies our empirical parametrization of the massloss history (eq. 14).  
Generally models with $n=1$ resemble a cored isothermal model. 
Models with $n \gg 1$ give a solid-body circular velocity curve, 
and models with $n \ll 1$ give a Keplerian-like curve with a dense solid-body core. 

If the parameters of the LMC are typical for satellites, then 
from the initial specific angular momentum $j_i \sim j_{\rm LMC} \sim 50\kpc \times 250\kms$,
we find the initial tidal radius $r_t$ is around 10 kpc (Fig.4b). 
The remnant is delivered to between 30-60 kpc depending
on $n$ if we also fix $p=1/3$.  

In order to deliver a final remnant to the inner 15 kpc,
we must start with an orbit with $j_i<j_{\rm LMC}$, hence the initial tidal radius of the satellite 
$r_t<10$kpc.  Fig.5 shows the range of predictions by varying the parameters $p$ and $n$.
The predicted circular velocity curves resemble those of the observed dwarf galaxies (Fig.4a),
but have typically high amplitudes with typically dense cores.  
Observed dwarf galaxies have cores of different densities (Fig.4a). 
The denser is the core of a dwarf galaxy, the further inward
is the final disrupted stream, but the remnants are never delivered 
to the very center.  Our models suggest that the remnants of most of dwarf galaxies 
should be typically outside the inner 15 kpc of the host galaxy halo.  

\section{Discussions}

\subsection{Effects of Disk and bulge}

There are two limitations of the current analysis.  
First we assume the effective mass is some simple function of time (cf. eq. 15),
whereas in reality several complex factors go into the modeling of
$m(t)$, the velocity function $\xi(u)$ and the Coulomb logarithm.
Second we assume an isothermal dark matter plus stars model throughout the galaxy, hence
the dynamical friction effect of the disk and bulge
are not modeled accurately.  However, these limitations should not 
change our qualitative conclusion that the remnants of outer halo satellites
should remain in the outer halo if tidal massloss is severe.  
For very eccentric orbits the factors in $\msat(t)$ may be modulated periodically
with the radius as the satellite oscillates between
pericenters and apocenters.  But these short-time-scale variations are smoothed out
and do not contribute to the evolution of the orbital angular momentum.  
In fact our results are very insensitive to the number of pericentric passages $N$
for $N$ between a few to a few hundred.
The disk and bulge are not important for our conclusion because 
we predict mainly the orbital decay in the outer halo where 
$j>j_{\rm disk}=3000{\rm kpc~km/s}$.  Inside 15 kpc, our estimation of dynamical friction 
by an SIS model is inaccurate only for satellites on low inclination orbits.
If satellites come in random inclinations, it is more common to find high inclination orbits,
for which our models should be fairly accurate even inside 15 kpc.

\subsection{Orbital decay of the progenitors of the Sgr and the Canis Major dwarf galaxies}

The Sgr dwarf and the Canis Major dwarf are the closest known dwarf galaxies, 
about 15 kpc from the center of the Milky Way and at the edge of the Milky Way disk.  
The Canis Major is on a (direct or retrograde) orbit 
slightly inclined from the plane of the Milky Way, and the Sgr is on a nearly polar orbit. 
Both orbits have a fairly low angular momentum with $S \sim 20\kpc$; the data on Sgr are 
more complete, and show that it oscillates between
10 kpc pericenter and 50 kpc apocenter.  Both contain 
several globular clusters.
It is possible that the two dwarfs are the stripped-down version of a more massive object,
which has dynamically decayed from the outer halo.  

Interestingly a possible 
extension of the Sgr has been reported recently in the SDSS data near the position of
the outer halo globular cluster NGC2419 (Newberg et al. 2003).
There is a stream-like enhancement of halo A-colored stars at the SDSS magnitude of $g_0=20.3$
in the plane of the Sgr's orbit, corresponding to a distance of 90 kpc.   
If this is true, it would imply that the Sgr
has changed its orbits in the past Hubble time.  There are two possible ways that
this could happen.  One is that the Sgr's orbit has been deflected by a massive
satellite, such as the LMC or SMC.  Indeed the orbits of the Sgr and the Magellanic Clouds
do overlap at the Galactic poles, and 
simple timing arguments show that these systems encounter or fly by
each other about 2.5 Gyrs ago at about 50 kpc on the North Galactic Pole if the rotation curve of the 
Milky Way is nearly flat (Zhao 1998).  The problem of this solution is that
it is rare for the Sgr to receive a strong enough deflection to bring down its orbit.

Another solution is that the Sgr has been a more massive system, which orbital decayed from
the outer halo (Jiang \& Binney 2000).  This solution requires a large amount of massloss.
Suppose that the progenitor of the Sgr has a mass comparable to the LMC, $10^{10}\msun$, and
was on an orbit with similar pericenter and apocenter as the LMC,
i.e., between 40 kpc and 100 kpc, then it would nicely explain the stream near NGC2419 as 
debris of the Sgr at an earlier time, when the Sgr was on a larger orbit with apocenter greater than 90 kpc 
(cf. Fig. 2).   

\subsection{The orbit of \omegacen and its progenitor}

We have mainly concentrated on the problem of getting rid of a satellite's angular momentum if it starts
with a high angular momentum or orbital size $S_0\gg 15$kpc.
What would be the remnant distribution if a satellite is born with an initial orbital size $S_i<15$kpc?  
The stars in such a system are assembled in the inner halo 
from the start, e.g., by colliding an infalling gas cloud with the protogalactic gas clouds in the inner halo, 
(Fellhauer \& Kroupa 2002).  Or the stars form from extragalactic gas and descend on a very radial orbit,
penetrating the inner 15 kpc of the host halo from its very first pericentric passage.  
An intriguing example is \omegacen.
Unfortunately our analytical model is not suited for this system because it is presently 
on a low-inclination eccentric retrograde orbits between 1 and 6 kpc from the Galactic center (Dinescu et al. 1999), 
so the contribution of dynamical friction by the disk is important.  Also hydrodynamical friction
with the disk gas can play a role for an early-on partially gaseous satellite.  Nevertheless, 
if one applies simplisticly the tidal massloss and ChandraSekhar's dynamical friction in a spherical halo,i
one finds that while it seems easy to peel off a satellite galaxy to make a central star cluster,
most simulations produce remnants on much larger orbits than \omegacen (Zhao 2002).  
It seems some fine tuning is required to select progenitors on very low angular momentum and/or low energy orbits:
the initial angular momentum needs to be low enough for 
the progenitor to penetrate into the inner halo or the present location of \omegacen on its 
very first pericentric passage.  This means that the initial orbital size $S_i^\omega$ of \omegacen is 
in between the present value of \omegacen $S_0 \sim 1.25$kpc and 
the boundary of the inner halo $R_{\rm disk}=15$kpc, or mathematically
\beq
1.25\kpc < S_i^\omega < 15\kpc.
\eeq

Most recently there have been several very encouraging attempts to model the dynamical and star formation history
of \omegacen by nearly self-consistent N-body simulations (Mizutani et al. 2003, Tsuchiya et al. 2003, Bekki \& Freeman 2003).
All are able to produce a reasonable mass and orbit of the \omegacen simultaneously after some trial and error 
with the initial parameters of the progenitor; 
many initial conditions lead to remnants, unlike \omegacen, beyond 10kpc of the Milky Way center. 
The favored initial orbit has a small orbital size $S_i$.
according to Tsuchiya et al. $j_i=60 {\rm kpc} \times 20 \kms =1200{\rm kpc}\kms$ (or $S_i=6$kpc) 
and  according to Bekki \& Freeman $j_i=25 {\rm kpc} \times 60 \kms =1500{\rm kpc}\kms$ (or $S_i=7.5$kpc).  
The small orbital size seems consistent with our expectation (cf. Fig.3).

Tsuchiya et al. launch satellites with various initial mass $(0.4-1.6)\times 10^{10}\msun$ 
and with either a King profile or a Hernquist profile from 60 kpc from the Milky Way center. 
They choose well-aimed nearly radial orbits, with an initial perigalactic radius about 1 kpc, 
much more radial than the present eccentric orbit.  
Massloss in their King model are similar to our exponential massloss models ($n=\infty$):
rapid in the beginning, and $\log(m)$ is roughly linear with time up to a mass of $10^8\msun$
when the satellite has too little mass to proceed with the orbital decay.
Massloss in the Hernquist model is closer to a $n=0.3$ model, linear in the beginning and 
rapid just before complete disruption. 

The progenitor in Tsuchiya et al.'s best simulation
is a two-component "nucleuated" model with 
a satellite of $0.8\times 10^{10}\msun$ with a Hernquist profile of half-mass radius $1.4$kpc 
plus a nucleus modeled by an extended-particle of $10^7\msun$ with a half mass radius of 35pc.  
A closer examination of this nucleus, however, reveals a subtle internally inconsistency
in the initial condition.  The nucleus is too fluffy
to be there in the first place.  Inside the half-mass radius of 35 pc of the nucleus, 
the nucleus or extended particle has less enclosed mass 
$M_{\rm nuc}(r)$ than the Hernquist main body of the satellite $M_{\rm Hern}(r)$,
or mathematically,
\beq
M_{\rm nucl}(r)=0.5 \times 10^7\msun \ll 
M_{Hern}(r)= 8 \times 10^9\msun \left({r \over r+(\sqrt{2}-1)\cdot 1414{\rm pc}}\right)^2 
=2\times 10^7, ~ r=35{\rm pc}.   
\eeq
So the "nucleus" is not a real self-gravitating identity, and should have been disintegrated by 
the tides of the Hernquist component if it was not kept as a rigid particle by construction.
If this extended particle were removed from the simulation, there will be no \omegacen-like remnant
left once the whole Hernquist satellite is disrupted by Galactic tides.  
One could stablise the initial condition by replacing the extended particle by a black hole or
a dense cluster with half-radius less than 35pc, but then the size of \omegacen is not reproduced
satisfactorily.  In this aspect, the nucleated dwarf models of Bekki \& Freeman are more self-consistent.

\section{Summary}

We have explored the orbital decay for a dwarf satellite with a range of
initial specific angular momentum and massloss history.  
We follow the evolution of satellites in the mass-distance plane, and find 
generally very little evolution of specific angular momentum by dynamical friction.
The progenitors of inner halo globular clusters and substructures 
can not be born on orbits of comparable angular momentum as present-day halo satellite 
galaxies. In general,
G1-type of objects at relatively large radius (assume G1 is on a circular orbit)
should be much more common than the \omegacen-type inner galaxy objects
if tidal stripping of outer halo dwarf satellites has been important in the
formation of massive galaxies.  It is also difficult for a minor merger to bring
a massive BH from outside the host galaxy all the way to the host galaxy center.  Satellite remnants
(BHs, globulars and streams) tend to hang up in the outer halo.

\acknowledgments 

I thank Ken Freeman, George Meylan, Jim Pringle, Floor van Leeuwen for enlightening discussions during the \omegacen conference at IoA,
and Oleg Gnedin, Mike Irwin, Pavel Kroupa, Mark Wilkinson for a careful reading of an earlier draft.
Special thanks to MianMian and YiYi (http://www.ast.cam.ac.uk/~hsz/Twins/FirstYear.html) 
for not making excessive noise and comments during this work.

\begin{figure*}
\epsfxsize=14cm
\epsfbox{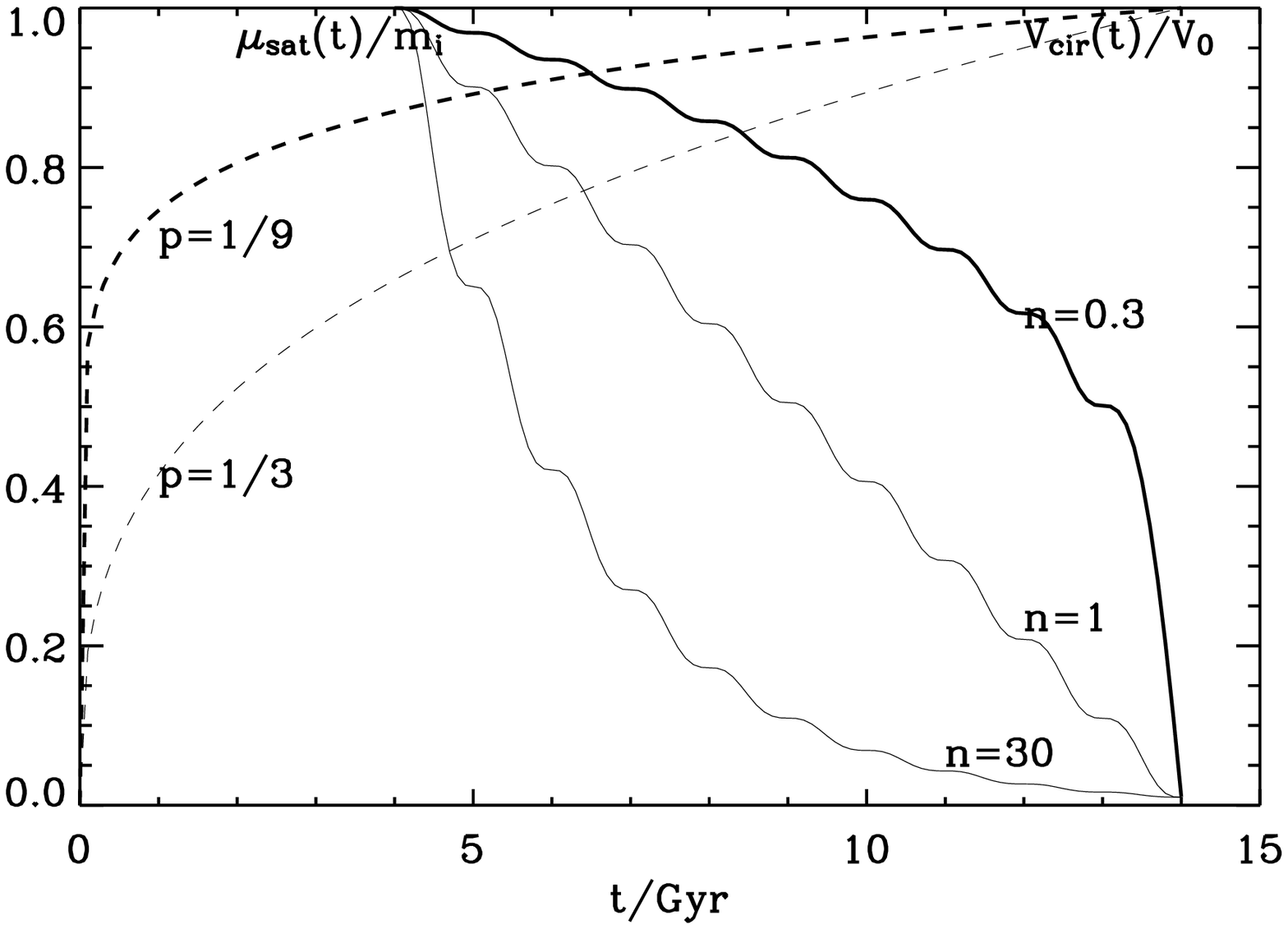}
\caption{
shows the satellite mass $\msat(t)$ as a function of time after rescaling by the initial mass $m_i$ 
(from top to bottom $n=0.3,1,\infty$),
and the rotation speed $V_{\rm cir}(t)$ (for top to bottom 
$p={1 \over 9},{1 \over 3}$) for the past Hubble time after rescaling by the present rotation speed $V_0$. 
}
\end{figure*}

\begin{figure*}
\epsfxsize=14cm
\epsfbox{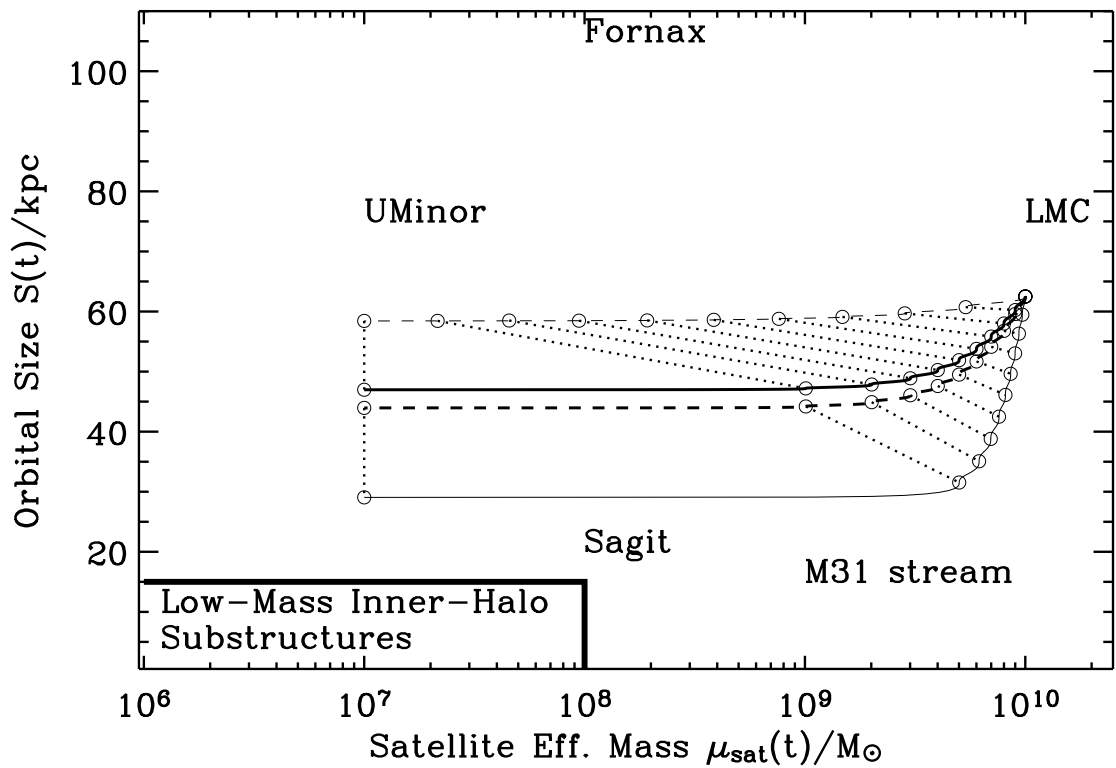}
\caption{
shows the predicted evolution histories of a satellite
in the plane of its effective mass $\msat(t)$ and  
characteristic orbital size $S(t)$ (which is the specific angular momentum $j(t)$ divided
by a chacteristic velocity $200\kms$ irrespective of eccentricity of the orbit).
Also indicated are the estimated orbital size 
and the mass of the satellite galaxies of the Milky Way and M31.
The upper hatched horn-like area shows models
with a massloss between exponential (upper dashed boundary, $n=\infty$) and linear 
(lower solid boundary, $n=1$); the potential grows moderately with $p={1 \over 3}$.  
The lower hatched horn-like area shows the predictions with a minor evolution of the potential ($p={1 \over 9}$)
with a massloss between linear (upper dashed boundary, $n=1$) and accelerated (lower solid boundary $n=0.3$).
A massive satellite starts at $t_i=4$ Gyrs from the upper right corner with 
an effective mass $\mu(t_i)=m_i=10^{10}\msun$ 
with angular momentum $j_i=250\kms\times 50$kpc, and ends with a mass of $10^7\msun$.
For different assumptions of the massloss rate,  
the intermediate mass and position of the remnant are indicated
with a time step of 1 Gyr.  Note the failure to deliver remnants to the lower left corner.     
}
\end{figure*}

\begin{figure*}
\epsfxsize=14cm
\epsfbox{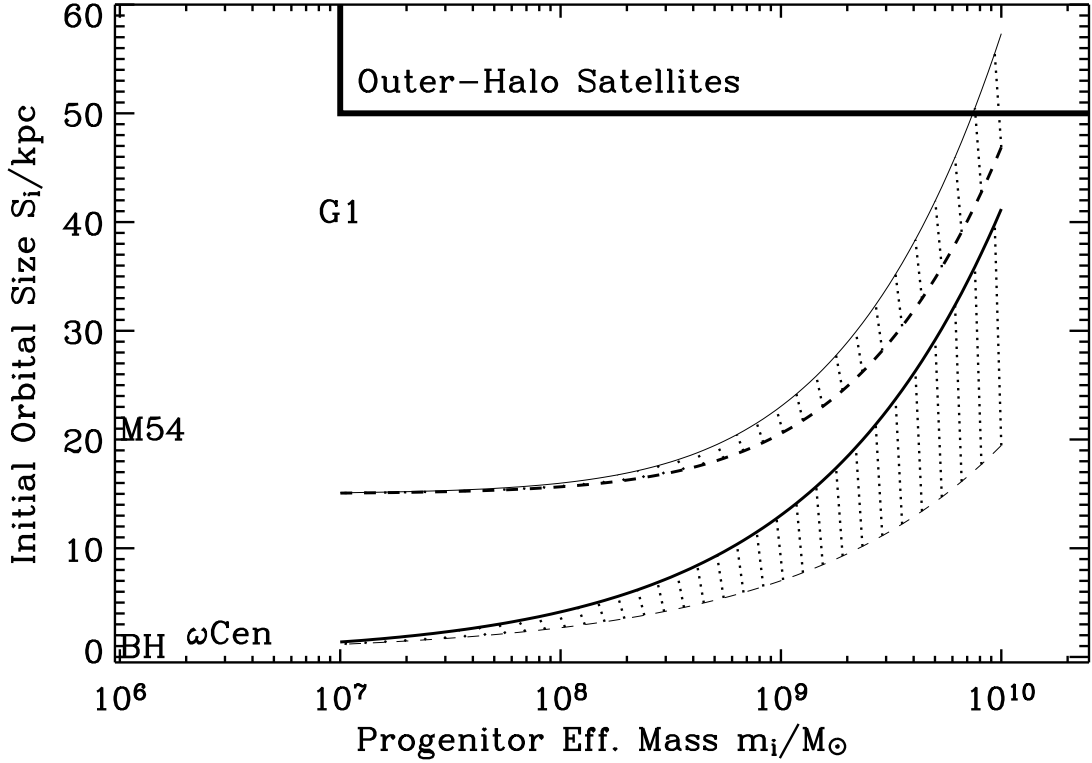}
\caption{
shows the initial orbital size $S_i=j_i/200$ and the progenitor mass of a satellite 
to deliver a small remnant ($10^6\msun$) to the inner galaxy (upper curves) and to the Galactic center (lower curves).  
An object in the upper left corner 
{\it cannot} evolve to a globular in the inner halo within a Hubble time. 
Different line types and shaded regions have the same meaning as in Figure 2.
Also indicated are the present values for the most massive globular clusters of the Milky Way and M31 and
a central million solar mass BH.
}
\end{figure*}

\begin{figure*}
\epsfxsize=10cm
\epsfbox{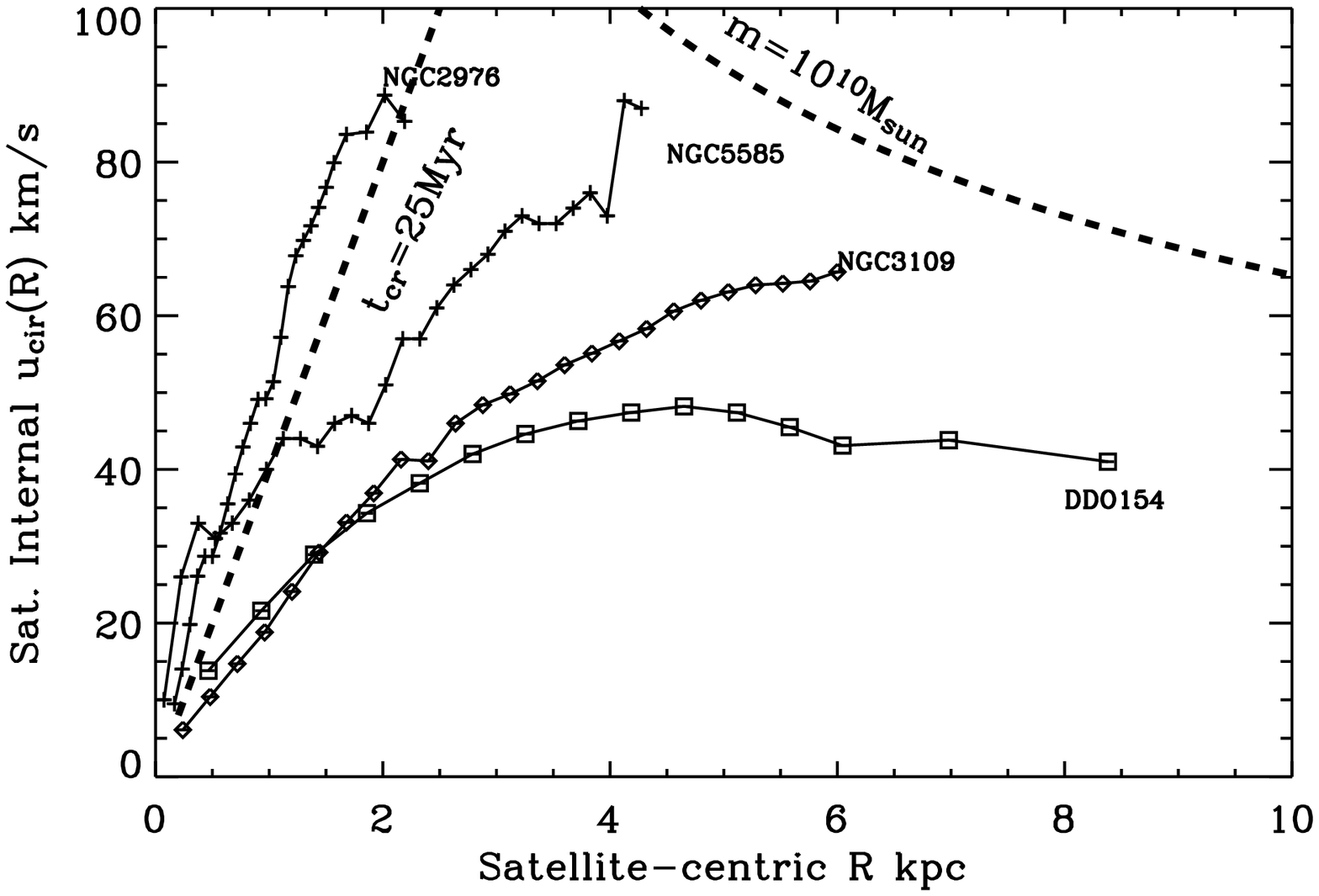}
\epsfxsize=10cm
\epsfbox{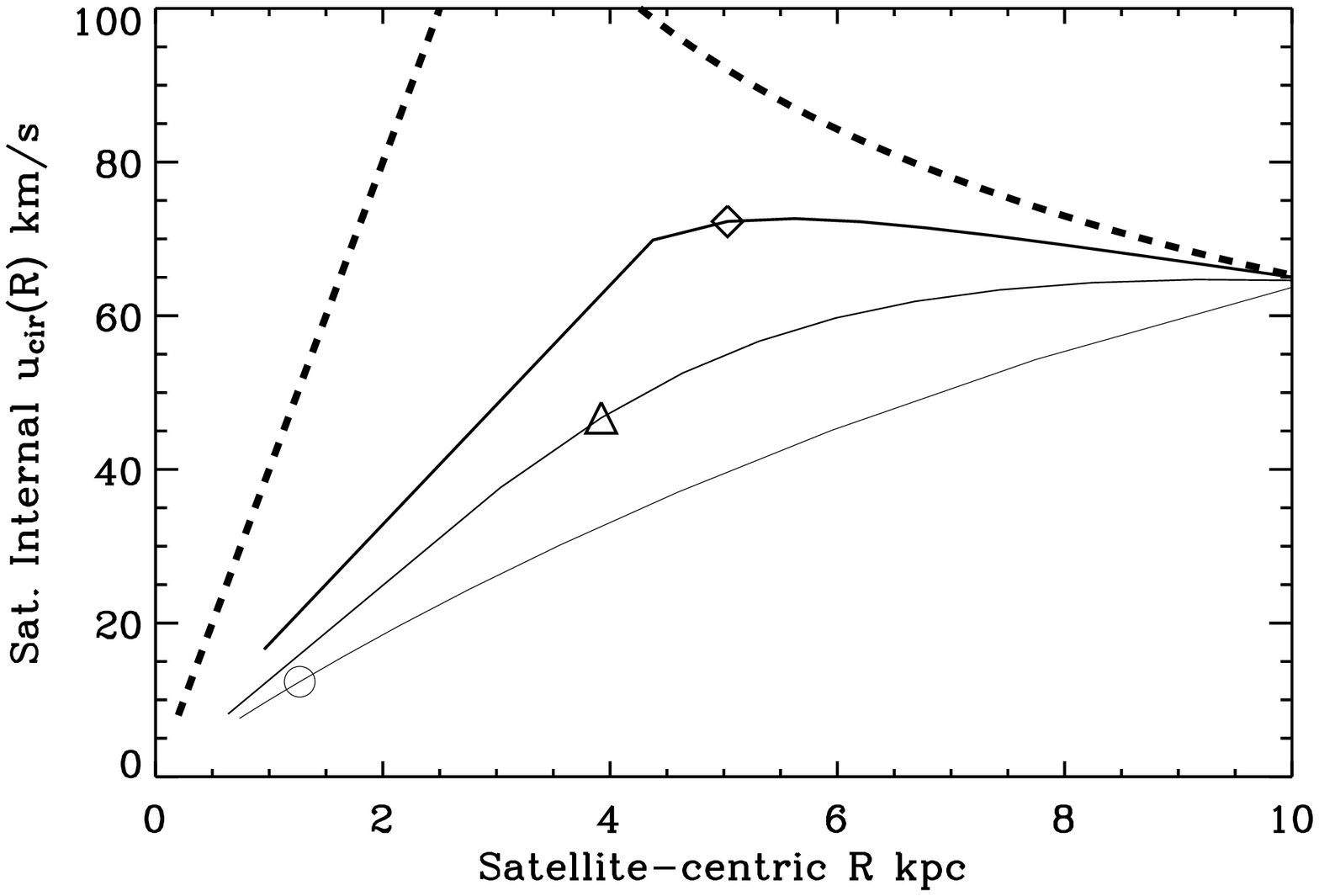}
\caption{compares the circular velocity curves 
of observed dwarf galaxies ({\bf upper panel}): DDO154 (Carnigan \& Purton 1998), NGC3109 (Jobin \& Carignan 1990), 
NGC5585 (Blais-Ouellette et al. 1999), NGC2976 (Simon et al. 2003)
with those infered from our model of massloss and dynamical friction ({\bf lower panel}).
The dashed thick line to the upper right is the Keplerian circular velocity curve of a point mass
of $10^{10}\msun$, to the lower left is solid body curve with a crossing time 25 Myr.
The three thinner curves are models with $n=0.3, 1, 30$ (up to down, marked in different symbols).
All models start with an initial mass $m_i=10^{10}\msun$ and 
on an eccentric orbit with peri-to-apo ratio $1:5$ 
and $j_i=50\kpc \times 250\kms$, similar to the LMC.   
The host galaxy has a moderate growth of the potential well with $p={1\over 3}$.  
All models have solid-body cores, each is labeled with the 
final orbital radius $S_0$ of the fully disrupted satellite (30kpc, 45 kpc, 60kpc from
up to down).
}
\end{figure*}

\begin{figure*}
\epsfxsize=14cm
\epsfbox{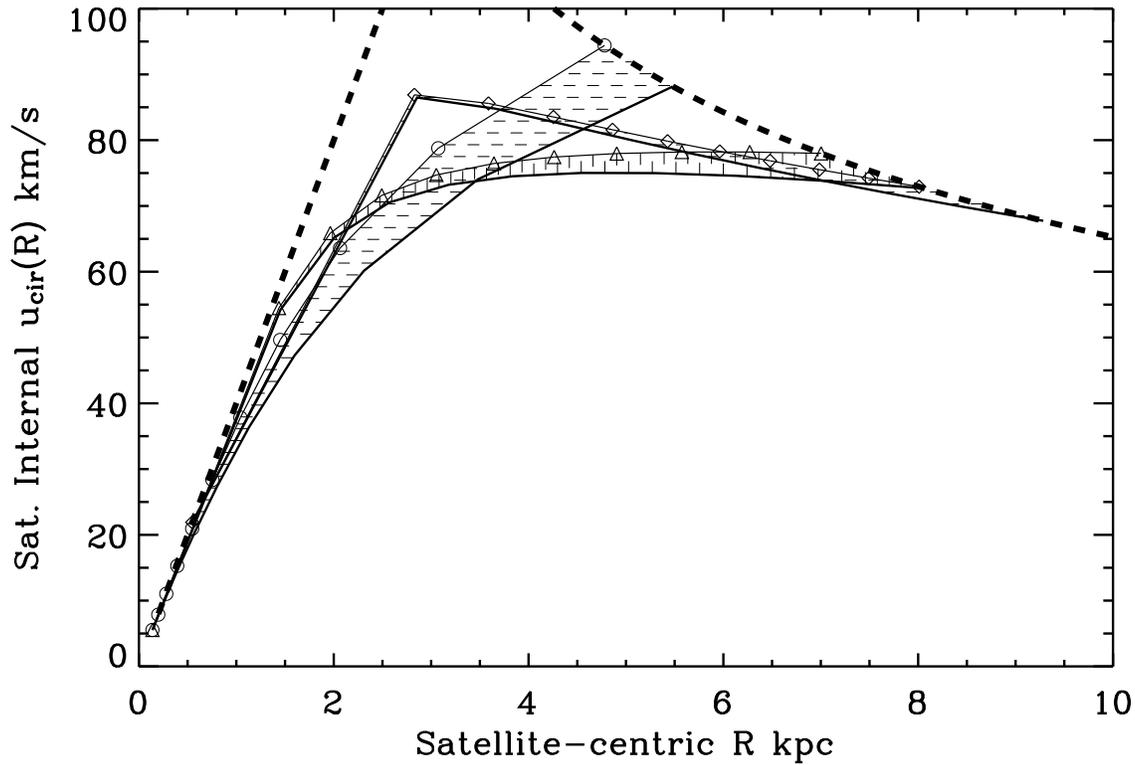}
\caption{Similar to Fig.4b with the three shades zones being models with $n=0.3, 1, 30$ 
and the upper and lower boundary of each zone being host halo models with $p={1 \over 9}$ and $p={1\over 3}$.  
All models end with an angular momentum $j_0=15\kpc \times 200\kms$ with pericenter outside the inner 15 kpc of the halo.
}
\end{figure*}

\vfill\eject

\end{document}